\documentclass[aps,prd,preprint]{revtex4}
\usepackage{graphicx}  
\usepackage{float}
\usepackage{bm}        
\usepackage{amssymb}   
\usepackage{slashed}   
\usepackage{amsmath}   
\usepackage{verbatim}  
\usepackage{color} 
\usepackage{xcolor} 
\usepackage{subfig} 
\usepackage{ulem}
\definecolor{Blue}{rgb}{0, 0.1, 0.5}

\usepackage[bookmarks=true,colorlinks=true,linkcolor=blue,unicode=true]{hyperref}
\graphicspath{{figure/}}
\newcounter{JW}

\begin{document}

\title{Constraints on anomalous quartic gauge couplings by $\gamma\gamma \to W^+W^-$ scattering}
\author{Yu-Chen Guo}
\email{ycguo@lnnu.edu.cn}
\author{Ying-Ying Wang}
\author{Ji-Chong Yang}
\email{yangjichong@lnnu.edu.cn}
\affiliation{Department of Physics, Liaoning Normal University, Dalian 116029, China}

\begin{abstract}
The vector boson scattering~(VBS) processes in Large Hadron Collider~(LHC) experiments offer a unique opportunity to probe the anomalous quartic gauge couplings~(aQGCs).
We study the dimension-8 operators contributing to the anomalous $\gamma\gamma WW$ coupling and the corresponding unitarity bounds via the exclusive $\gamma\gamma \to W^+W^-$ production in $pp$ collisions at LHC for a center of mass energy of $\sqrt{s}=13$ TeV.
By analysing the kinematical features of the signal, we propose an event selection strategy to highlight the aQGC contributions.
Based on the event selection strategy, the statistical significance of the signals are analyzed in detail, and the constraints on the coefficients of the anomalous quartic gauge operators are obtained.

\end{abstract}

\maketitle

\section{\label{level1}Introduction}

After the discovery of the Higgs boson~\cite{HiggsDiscover}, lots of effort has been made in the search of new physics beyond the Standard Model~(BSM) at the Large Hadron Collider~(LHC).
Among the abundant processes accessible in the LHC, the vector boson scattering~(VBS) processes draw a lot of attention because they are sensitive to BSM effects~\cite{VBSReview,VBSReview2}.
In the Standard Model~(SM), because of the cancellation among the VBS Feynman diagrams, the cross-sections of VBS do not grow with energy.
Such cancellation will be broken when BSM effects are involved~\cite{VBSNP}, therefor the VBS processes at high energies provide great opportunities to search for new physics~\cite{wwaa}.

In the search of BSM, the Standard Model Effective Field Theory (SMEFT)~\cite{SMEFTReview} has emerged as a powerful model-independent approach.
In this approach, after integrating out the BSM degree of freedom, the effects of BSM become higher dimensional operators suppressed by energy scale $\Lambda$, and the effective Lagrangian is
\begin{equation}
\begin{split}
&\mathcal{L}_{\rm SMEFT}=\mathcal{L}_{SM}+\sum _i\frac{C_{6i}}{\Lambda^2}\mathcal{O}_{6i}+\sum _j\frac{C_{8j}}{\Lambda^4}\mathcal{O}_{8j}+\ldots,
\end{split}
\label{eq.1.1}
\end{equation}
where $\mathcal{O}_{6i}$ and $\mathcal{O}_{8i}$ are dimension-6 and dimension-8 operators.
Odd dimension operators are neglected in this letter.
The high dimensional operators can contribute to the anomalous trilinear gauge boson couplings~(aTGCs) and anomalous quartic gauge boson couplings~(aQGCs) which are suitable to be studied via the VBS processes.

In this work, we focus on the aQGCs because the aTGC signals are sensitive else where, e.g. in diboson production processes and vector boson fusion processes~\cite{VBSNP}.
Moreover, we consider only the dimension-8 operators contributing to the aQGCs because the dimension-8 operators can introduce decorrelation between aTGCs and aQGCs, i.e. the dimension-6 operators can not contribute to QGCs while not affecting TGCs~\cite{VBSReview}.
There are also cases where the dimension-6 operators are absent and the dimension-8 operators are presented. For example,
the Born-Infeld~(BI) model~\cite{BIModel} with the Lagrangian
\begin{equation}
\begin{split}
&\mathcal{L}_{BI}=\beta^2\left[1-\sqrt{1+\sum _{i}^{12}\frac{F^{i,\mu\nu}F^i_{\mu\nu}}{2\beta^2}-\left(\sum _{i}^{12}\frac{F^{i,\mu\nu}\tilde{F}^i_{\mu\nu}}{4\beta^2}\right)^2}\right],\\
\end{split}
\label{eq.1.2}
\end{equation}
where $i$ corresponds to one of the $12$ generators of the SM $SU(3)_c\otimes SU(2)_L\otimes U(1)_Y$ gauge group, and $\tilde{F}_{\mu\nu}=\epsilon^{\mu\nu\alpha\beta}F_{\alpha\beta}/2$.

The evidence of the same sign $WWjj$ channel was found at LHC in 2014~\cite{SameSignWW}, which is the first evidence of the processes involving a QGC. Shortly afterwards, the dimension-8 operators contributing to aQGCs were studied in the VBS processes extensively, for example, in the same sign $WWjj$ channel~\cite{SameSignWWaQGC,SameSignWWAndMolCut}, $\gamma Wjj$ channel~\cite{awaw8TeV}, $ZZjj$ channel~\cite{zzjj13TeV}, $WZjj$ channel~\cite{WZ13TeV} and also $W^+W^-jj$ channel at $\sqrt{s}=7$ and $8\ {\rm TeV}$~\cite{aaWW8TeV}.
The evidence of exclusive or quasi-exclusive $\gamma \gamma \to W^+W^-$ process has been observed recently~\cite{aaWW8TeV}.
As a supplementary, we study aQGCs in this process at $\sqrt{s}=13\;{\rm TeV}$.
The $W^+W^-jj$ channel receive contributions from $\gamma\gamma WW$, $\gamma ZWW$, $ZZWW$ and $WWWW$ vertices~\cite{aQGCOperatorNew}, and one cannot discriminate those vertices by $W^+W^-jj$ channel alone.
Thus, we only consider the vertices contributing to the exclusive $\gamma\gamma\to W^+W^-$ process in this work.

It is well known that the rapid growth of the scattering amplitudes with energy leads to unitarity violation~\cite{UnitarityHistory}.
In this work, we calculate the partial wave unitarity bounds of the aQGCs at various proton c.m. energies.
We also study the kinematical features of the aQGC signal and the corresponding backgrounds.
The signal is found to be sensitive to the $M_{ol}$ cut which is used in the same sign $WWjj$ channel~\cite{SameSignWWAndMolCut}.
Except for that, the signal has a unique $\cos (\theta _{ll})$ behaviour which provides an efficient cut.
Based on the Monte Carlo~(MC) simulation, we estimate the constraints and observability of the anomalous $\gamma\gamma WW$ couplings with the current luminosity of LHC.

\section{\label{level2}The dimension-8 anomalous quartic gauge operators and $\gamma \gamma WW$ vertex}

We follow Refs.~\cite{aQGCOperatorNew,aQGCOperatorOld} to list all dimension-8 operators contributing to aQGCs, they are
\begin{equation}
\begin{split}
&\mathcal{L}_{aQGC}=\sum _{i=0}^2 \frac{f_{S_i}}{\Lambda^4}O_{S_i}+\sum _{j=0}^7 \frac{f_{M_j}}{\Lambda^4}O_{M_j}+\sum _{k=0}^9 \frac{f_{T_k}}{\Lambda^4}O_{T_k},
\end{split}
\label{eq.2.1}
\end{equation}
with
\begin{equation}
\begin{split}
&O_{S_0}=\left[\left(D_{\mu}\Phi \right) ^{\dagger} D_{\nu}\Phi\right]\times \left[\left(D^{\mu}\Phi \right) ^{\dagger} D^{\nu}\Phi\right],
 O_{S_1}=\left[\left(D_{\mu}\Phi \right) ^{\dagger} D_{\mu}\Phi\right]\times \left[\left(D^{\nu}\Phi \right) ^{\dagger} D^{\nu}\Phi\right],\\
&O_{S_2}=\left[\left(D_{\mu}\Phi \right) ^{\dagger} D_{\nu}\Phi\right]\times \left[\left(D^{\nu}\Phi \right) ^{\dagger} D^{\mu}\Phi\right],\\
\end{split}
\label{eq.2.2}
\end{equation}
\begin{equation}
\begin{array}{ll}
O_{M_0}={\rm Tr\left[\widehat{W}_{\mu\nu}\widehat{W}^{\mu\nu}\right]}\times \left[\left(D^{\beta}\Phi \right) ^{\dagger} D^{\beta}\Phi\right],
&O_{M_1}={\rm Tr\left[\widehat{W}_{\mu\nu}\widehat{W}^{\nu\beta}\right]}\times \left[\left(D^{\beta}\Phi \right) ^{\dagger} D^{\mu}\Phi\right],\\
O_{M_2}=\left[B_{\mu\nu}B^{\mu\nu}\right]\times \left[\left(D^{\beta}\Phi \right) ^{\dagger} D^{\beta}\Phi\right],
&O_{M_3}=\left[B_{\mu\nu}B^{\nu\beta}\right]\times \left[\left(D^{\beta}\Phi \right) ^{\dagger} D^{\mu}\Phi\right],\\
O_{M_4}=\left[\left(D_{\mu}\Phi \right)^{\dagger}\widehat{W}_{\beta\nu} D^{\mu}\Phi\right]\times B^{\beta\nu},
&O_{M_5}=\left[\left(D_{\mu}\Phi \right)^{\dagger}\widehat{W}_{\beta\nu} D_{\nu}\Phi\right]\times B^{\beta\mu} + h.c.,\\
O_{M_7}=\left(D_{\mu}\Phi \right)^{\dagger}\widehat{W}_{\beta\nu}\widehat{W}_{\beta\mu} D_{\nu}\Phi,&\\
\end{array}
\label{eq.2.3}
\end{equation}
\begin{equation}
\begin{array}{ll}
O_{T_0}={\rm Tr}\left[\widehat{W}_{\mu\nu}\widehat{W}^{\mu\nu}\right]\times {\rm Tr}\left[\widehat{W}_{\alpha\beta}\widehat{W}^{\alpha\beta}\right],
&O_{T_1}={\rm Tr}\left[\widehat{W}_{\alpha\nu}\widehat{W}^{\mu\beta}\right]\times {\rm Tr}\left[\widehat{W}_{\mu\beta}\widehat{W}^{\alpha\nu}\right],\\
O_{T_2}={\rm Tr}\left[\widehat{W}_{\alpha\mu}\widehat{W}^{\mu\beta}\right]\times {\rm Tr}\left[\widehat{W}_{\beta\nu}\widehat{W}^{\nu\alpha}\right],
&O_{T_5}={\rm Tr}\left[\widehat{W}_{\mu\nu}\widehat{W}^{\mu\nu}\right]\times B_{\alpha\beta}B^{\alpha\beta},\\
O_{T_6}={\rm Tr}\left[\widehat{W}_{\alpha\nu}\widehat{W}^{\mu\beta}\right]\times B_{\mu\beta}B^{\alpha\nu},
&O_{T_7}={\rm Tr}\left[\widehat{W}_{\alpha\mu}\widehat{W}^{\mu\beta}\right]\times B_{\beta\nu}B^{\nu\alpha},\\
O_{T_8}=B_{\mu\nu}B^{\mu\nu}\times B_{\alpha\beta}B^{\alpha\beta},
&O_{T_9}=B_{\alpha\mu}B^{\mu\beta}\times B_{\beta\nu}B^{\nu\alpha},\\
\end{array}
\label{eq.2.4}
\end{equation}

where $\widehat{W}\equiv \vec{\sigma}\cdot {\vec W}/2$ with $\sigma$ the Pauli matrix and ${\vec W}=\{W^1,W^2,W^3\}$. Note that, we keep the index of the operators identical to Ref.~\cite{aQGCOperatorOld}, and therefor the redundant ($O_{M_6}$) or vanishing operators ($O_{T_{3,4}}$) are not included. The operators contributing to the $\gamma\gamma WW$ interaction can form $5$ different vertices $\mathcal{L}_{AAWW}=\sum _{i=0}^4 \alpha_i V_{i,AW}$, where
\begin{equation}
\begin{array}{ll}
V_{0,AW}=F_{\mu\nu}F^{\mu\nu}W^{+\alpha}W^-_{\alpha}, &V_{1,AW}=F_{\mu\nu}F^{\mu\alpha}W^{+\nu}W^-_{\alpha}\\
V_{2,AW}=F_{\mu\nu}F^{\mu\nu}W^+_{\alpha\beta}W^{-\alpha\beta}, &V_{3,AW}=F_{\mu\nu}F^{\nu\alpha}W^+_{\alpha\beta}W^{-\beta\mu}\\
V_{4,AW}=F_{\mu\nu}F^{\alpha\beta}W^+_{\mu\nu}W^{-\alpha\beta},& \\
\end{array}
\label{eq.2.6}
\end{equation}
where $W^{\pm\mu\nu}\equiv \partial _{\mu}W^{\pm}_{\nu}-\partial _{\nu}W^{\pm}_{\mu}$. The coefficients are
\begin{equation}
\begin{split}
&\alpha_0=\frac{e^2v^2}{8\Lambda ^4}\left(f_{M_0}+\frac{c_W}{s_W}f_{M_4}+2\frac{c_W^2}{s_W^2}f_{M_2}\right),
\alpha_1=\frac{e^2v^2}{8\Lambda ^4}\left(\frac{1}{2}f_{M_7}+2\frac{c_W}{s_W}f_{M_5}-f_{M_1}-2\frac{c_W^2}{s_W^2}f_{M_3}\right),\\
&\alpha_2=\frac{1}{\Lambda ^4}\left(s_W^2f_{T_0}+c_W^2f_{T_5}\right),\;
 \alpha_3=\frac{1}{\Lambda ^4}\left(s_W^2f_{T_2}+c_W^2f_{T_7}\right),\;
 \alpha_4=\frac{1}{\Lambda ^4}\left(s_W^2f_{T_1}+c_W^2f_{T_6}\right).\\
\end{split}
\label{eq.2.7}
\end{equation}
$V_{i=0,1, AW}$ are dimension-6 vertices, and $V_{i=2,3,4, AW}$ are dimension-8 vertices which can be introduced by BI model.

One dimension-8 operator contribute to only one vertex, therefor the constraints on $\alpha _i$ can be derived by the constraints on dimension-8 operators.
The range of $\alpha _i$ depends on maximum of each $f$ term.
Base on the experimental limits of $f_{M}$ and $f_{T}$~\cite{awaw8TeV}, we get the tightest constraints in Table.~\ref{tab.1}.
\begin{table}
\caption{\label{tab.1}The constraints on vertices and the corresponding limits on the dimension-8 operators.
}
\centering
\begin{tabular}{cc|cc}
\hline
vertex & constraint & coefficient & constraint\\
\hline
$\alpha_0 ({\rm TeV^{-2}})$ & $[-0.12, 0.12]$ & $f_{M_2}/\Lambda ^4\;({\rm TeV^{-4}})$ & $[-26, 26]$ \\
$\alpha_1 ({\rm TeV^{-2}})$ & $[-0.20, 0.20]$ & $f_{M_3}/\Lambda ^4\;({\rm TeV^{-4}})$ & $[-43, 44]$ \\
$\alpha_2 ({\rm TeV^{-4}})$ & $[-2.9, 2.9]$ & $f_{T_5}/\Lambda ^4\;({\rm TeV^{-4}})$ & $[-3.8, 3.8]$ \\
$\alpha_3 ({\rm TeV^{-4}})$ & $[-5.9, 5.9]$ & $f_{T_7}/\Lambda ^4\;({\rm TeV^{-4}})$ & $[-7.3, 7.7]$ \\
$\alpha_4 ({\rm TeV^{-4}})$ & $[-2.3, 2.3]$ & $f_{T_6}/\Lambda ^4\;({\rm TeV^{-4}})$ & $[-2.8, 3.0]$ \\
\hline
\end{tabular}
\end{table}

\section{\label{level3} Unitarity bound}

The aQGC contributions grow significantly at high energies.
On one hand, this feature indicates that at higher energies, the VBS process is ideal to search for aQGCs, on the other hand.
The cross-section of VBS with aQGCs will violate unitarity at certain energy, which indicates that as the energy scale grows the new physics particles degrees of freedom will emerge, and SMEFT is not valid.
To avoid the violation of unitarity, the coefficients of the operators will be constrained, which is the unitarity bound.

Considering the process $\gamma_{\lambda _1}\gamma_{\lambda _2}\to W^-_{\lambda _3}W^+_{\lambda _4}$, where $\lambda _{1,2}=\pm 1$ and $\lambda _{3,4}=\pm 1, 0$ correspond to the helicity of the vector bosons, the amplitudes can be expanded as~\cite{PartialWave,UnitaryBound}
\begin{equation}
\begin{split}
&\mathcal{M}(\gamma_{\lambda _1}\gamma_{\lambda _2}\to W^-_{\lambda _3}W^+_{\lambda _4})=8\pi \sum _{J}\left(2J+1\right)\sqrt{1+\delta _{\lambda _1\lambda _2}}\sqrt{1+\delta _{\lambda _3\lambda _4}}e^{i(\lambda-\lambda ') \varphi}d^J_{\lambda \lambda '}(\theta) T^J\\
\end{split}
\end{equation}
where $\lambda = \lambda _1-\lambda _2$, $\lambda ' =\lambda _3-\lambda _4$ and $d^J_{\lambda \lambda '}(\theta)$ is the Wigner $d$-function~\cite{PartialWave}.
The partial wave unitarity bound is $|T^J|\leq 2$~\cite{UnitaryBound} which is widely used~\cite{UnitaryBoundOthers}.

For the $\gamma \gamma \to W ^+W^-$ process, $36$ different helicity amplitudes can be obtained by partial wave expansion.
The number of amplitudes can be reduced by using $\mathcal{M}_{\lambda _1, \lambda _2, \lambda _3, \lambda _4}(\theta ) = (-1)^{\lambda _1-\lambda _2-\lambda _3+\lambda _4}\mathcal{M}_{-\lambda _1, -\lambda _2, -\lambda _3, -\lambda _4}(\theta )$.
For simplicity we denote $\hat{s}=(p_{\gamma 1}+p_{\gamma 2})^2$, note that $\hat{s}$ is not the c.m. energy of protons.
It is only necessary to keep the terms at the leading order $\mathcal{O}(\hat{s}^2)$, which grow fastest with $\hat{s}$. The leading terms are list in Table.~\ref{tab.Helicity}.
\begin{table}
\begin{center}
\begin{tabular}{c|ccccc}
\hline
Amplitudes & $\alpha _0$ & $\alpha _1$ & $\alpha _2$ & $\alpha _3$ & $\alpha _4$ \\
\hline
$|\mathcal{M}_{++++}|$ & $\mathcal{O}(\hat{s})$ & $\mathcal{O}(\hat{s})$ & $2\alpha _2 \hat{s}^2$ & $\frac{1}{2}\alpha _3 \hat{s}^2$ & $\mathcal{O}(\hat{s}^0)$ \\
$|\mathcal{M}_{++--}|$ & $\mathcal{O}(\hat{s})$ & $\mathcal{O}(\hat{s})$ & $2\alpha _2 \hat{s}^2$ & $\frac{1}{2}\alpha _3 \hat{s}^2$ & $\frac{1}{4}\alpha _4 \hat{s}^2 \left(\cos(2\theta)+3\right)$ \\
$|\mathcal{M}_{++00}|$ & $\alpha _0\frac{\hat{s}^2}{M_W^2}$ & $\frac{1}{4}\alpha _1\frac{\hat{s}^2}{M_W^2}$ & $\mathcal{O}(\hat{s})$ & $\mathcal{O}(\hat{s})$ & $\mathcal{O}(\hat{s})$ \\
$|\mathcal{M}_{+-+-}|$ & $\mathcal{O}(\hat{s}^0)$ & $\mathcal{O}(\hat{s})$ & $\mathcal{O}(\hat{s}^0)$ & $\frac{1}{2}\alpha _3 \hat{s}^2\cos^4(\frac{\theta}{2})$ & $\alpha _4 s^2\cos^4(\frac{\theta}{2})$ \\
$|\mathcal{M}_{+--+}|$ & $\mathcal{O}(\hat{s}^0)$ & $\mathcal{O}(\hat{s})$ & $\mathcal{O}(\hat{s}^0)$ & $\frac{1}{2}\alpha _3\hat{s}^2\sin^4(\frac{\theta}{2})$ & $\alpha _4\hat{s}^2\sin^4(\frac{\theta}{2})$ \\
$|\mathcal{M}_{+-00}|$ & $\mathcal{O}(\hat{s}^0)$ & $\frac{\alpha _1 \hat{s}^2 \sin ^2(\theta)}{8 M_W^2}$ & $\mathcal{O}(\hat{s}^0)$ & $\mathcal{O}(\hat{s})$ & $\mathcal{O}(\hat{s})$ \\
\hline
\end{tabular}
\end{center}
\caption{\label{tab.Helicity}The helicity ampltidues at the order of $\mathcal{O}(s^2)$.}
\end{table}
The tightest bounds are
\begin{equation}
\begin{split}
&\left|\alpha _0\right|\leq \frac{32\pi M_W^2}{\hat{s}^2},\;\;\left|\alpha _1\right|\leq \frac{128\pi M_W^2}{\hat{s}^2},\;\;
 \left|\alpha _2\right|\leq \frac{16\pi}{\hat{s}^2},\;\;\left|\alpha _3\right|\leq \frac{64\pi }{\hat{s}^2},\;\;\left|\alpha _4\right|\leq \frac{48\pi}{\hat{s}^2}.
\end{split}
\end{equation}
$\sqrt{\hat{s}}$ is related to $\sqrt{s}$ by photon distribution functions~\cite{PhotonDistrion}.
To extract $\sqrt{\hat{s}}$, we analysis the $\gamma \gamma \to l^+l^-\nu\bar{\nu}$ process with photons from protons based on Monte Carlo~(MC) simulation using the \verb"MadGraph5_aMC@NLO" toolkit~\cite{madgraph,madanalysis,pythia,delphes}.
We run MC simulations for $\sqrt{s}=13\ (14)$ TeV at LHC \cite{HL-LHC}, 27 TeV at HE-LHC~\cite{HE-LHC}, 50 TeV at FCC-hh~\cite{FCC-hh} and 100 TeV at SppC~\cite{SppC}.
We find that the energies of photons grow very slowly with $\sqrt{s}$.

Since $\sqrt{\hat{s}}$ is a distribution, one can set the unitarity bounds in a statistical way.
We set the bounds by requiring $95\%$ events are at the valid region in the sense of unitarity. 
The corresponding $\sqrt{\hat{s}}$ and the bounds on the coefficients are listed in Table.~\ref{tab.Unitary}.
It can be found that the ranges of the coefficients in Table.~\ref{tab.1} indeed satisfy the unitarity bounds.
\begin{table}
\begin{center}
\begin{tabular}{c|c|ccccc}
\hline
$\sqrt{s}$ & $\sqrt{\hat{s}}$ & $\alpha _0$ ($\rm TeV^{-2}$)& $\alpha _1$ ($\rm TeV^{-2}$) & $\alpha _2$ ($\rm TeV^{-4}$) & $\alpha _3$ ($\rm TeV^{-4}$) & $\alpha _4$ ($\rm TeV^{-4}$) \\
\hline
$13$ TeV & $0.78$ TeV & $|\alpha _0|<1.75$ & $|\alpha _1|<7.02$ & $|\alpha _2|<136$ & $|\alpha _3|<543$ & $|\alpha _4|<407$ \\
$14$ TeV & $0.79$ TeV & $|\alpha _0|<1.67$ & $|\alpha _1|<6.67$ & $|\alpha _2|<129$ & $|\alpha _3|<516$ & $|\alpha _4|<387$ \\
$27$ TeV & $0.90$ TeV & $|\alpha _0|<0.99$ & $|\alpha _1|<3.96$ & $|\alpha _2|<77$ & $|\alpha _3|<306$ & $|\alpha _4|<230$ \\
$50$ TeV & $0.99$ TeV & $|\alpha _0|<0.68$ & $|\alpha _1|<2.70$ & $|\alpha _2|<52$ & $|\alpha _3|<209$ & $|\alpha _4|<157$ \\
$100$ TeV & $1.07$ TeV & $|\alpha _0|<0.50$ & $|\alpha _1|<1.98$ & $|\alpha _2|<38$ & $|\alpha _3|<153$ & $|\alpha _4|<115$ \\
\hline
\end{tabular}
\end{center}
\caption{\label{tab.Unitary}The unitarity bounds on the vertices at different c.m. energies.}
\end{table}

\section{\label{level4} Statistical significance of anomalous quartic gauge operators}

\subsection{\label{level4.1} Signal and Background Analysis}

\begin{figure}
\begin{center}
\includegraphics[width=0.6\textwidth]{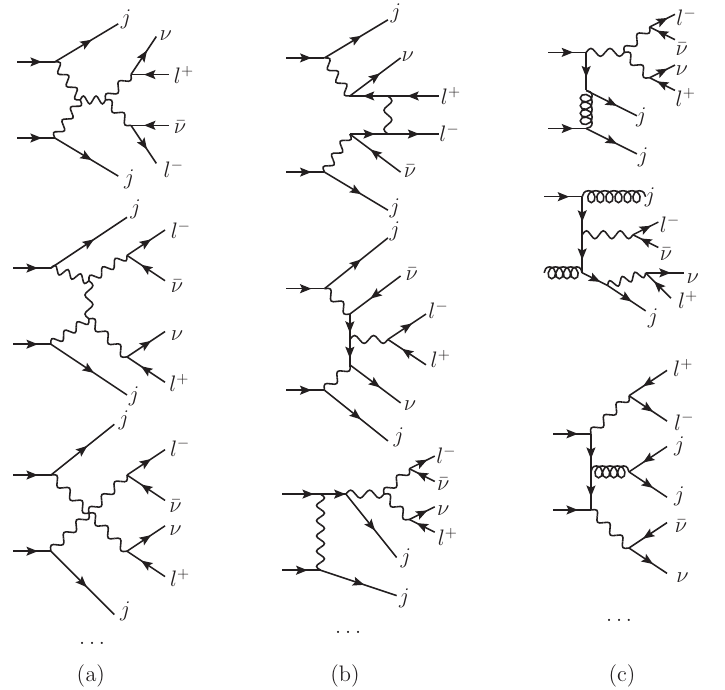}
\caption{The backgrounds are the processes contributing to $jjl^+l^-\nu\bar{\nu}$ final states in the SM. The typical EW-VBS diagrams are shown in (a), EW-non-VBS diagrams are shown in (b), and the typical QCD diagrams are shown in (c).}
\label{fig:typicalBackground}
\end{center}
\end{figure}

\begin{figure}
\begin{center}
\includegraphics[width=0.55\textwidth]{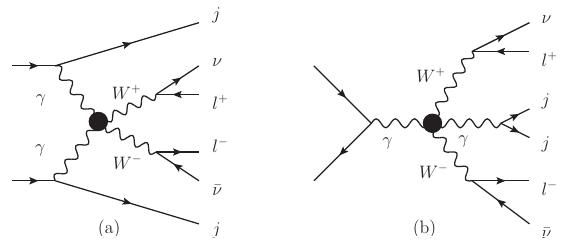}
\caption{The typical diagrams with anomalous $\gamma\gamma WW$ coupling contributing to $jj\ell^+\ell^-\nu\bar{\nu}$ final states. Similar as the SM, there are also VBS contributions and non-VBS contributions.}
\label{fig:typicalSignal}
\end{center}
\end{figure}

Experimentally, VBS is characterized by the presence of a pair of vector bosons and two forward jets.
The SM process $pp\to jj\ell^+\ell^-\nu\bar{\nu}$ is treated as the irreducible background, the typical Feynman diagrams at tree level are shown in Fig.~\ref{fig:typicalBackground}, which can be categorized into three groups, electromagnetic-weak VBS~(EW-VBS), electromagnetic-weak non-VBS~(EW-non-VBS), and QCD diagrams.
The signal is defined as the dimension-8 operators induced $\gamma \gamma\to W^+W^-$ process with leptonic decay, and we consider one vertex at a time.
The typical Feynman diagram of the signal is shown in Fig.~\ref{fig:typicalSignal}.~(a), while the $s$-channel diagrams (e.g. Fig.~\ref{fig:typicalSignal}.~(b)) are excluded because they do not contribute to $\gamma\gamma \to W^+W^-$ process.
In order to suppress the contribution of diagrams with a $Wtb$ vertex in electro-weak $W^+W^-jj$ production, especially the $t\bar{t}$ backgrounds, we exclude the $b$-jets in our events.
The signal events are generated with the largest values of $\alpha _i$ in Table.~\ref{tab.1}. For both the signals and the background a CMS-like detector simulation is applied using the \verb"Delphes" framework~\cite{delphes}. The analyses of the signals and the background are completed by \verb"MLAnalysis"~\cite{Guo:2023nfu}.  

\begin{figure}
\subfloat[$M_{jj}$]{\includegraphics[width=0.5\textwidth]{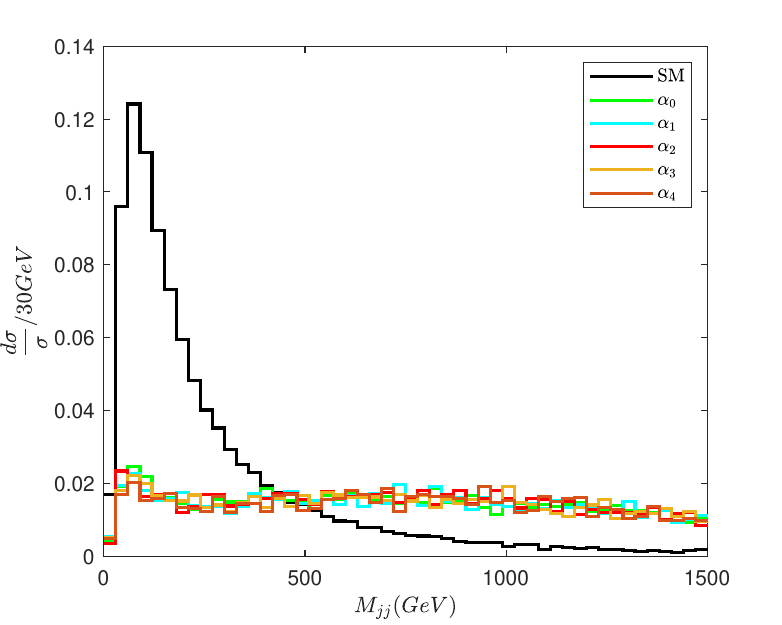}}\hfill
\subfloat[$|\Delta y_{jj}|$]{\includegraphics[width=0.5\textwidth]{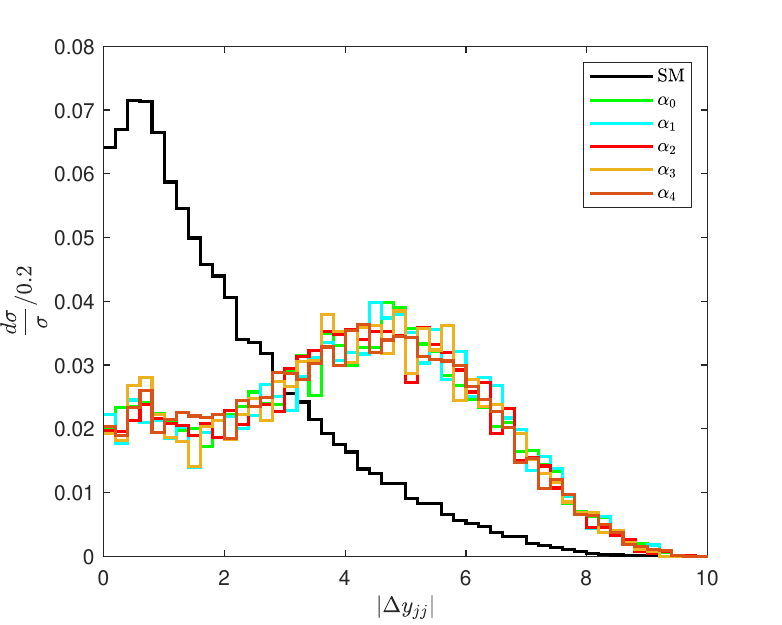}}\vfill
\caption{The differential cross-sections of the SM background and signals as a function of $M_{jj}$ (left panel) and $|\Delta y_{jj}|$ (right panel) after the $N_{\ell}$ and $N_j$ cuts.}
\label{fig:vbfcuts}
\end{figure}

Events are required to have exactly two opposite sign charged leptons (electrons or muons) with at least $2$ jets, i.e. $N_{\ell}=2$, $N_{j}\geq 2$.
To extract the VBS contribution, the standard VBF/VBS cuts are applied~\cite{VBFCut}.
The distributions for signals and background of the invariant dijet mass ($M_{jj}$) and rapidity separation ($|\Delta y_{jj}|$) are shown in Fig.~\ref{fig:vbfcuts}.
Since the cuts on the leptonic final states are very efficient, we choose relatively loose cuts $M_{jj}>180{\rm\ GeV}$ and $\left|\Delta y_{jj}\right|>2.3$ to increase the signal yield.

\begin{figure}
\subfloat[$M_{o1}$]{\includegraphics[width=0.5\textwidth]{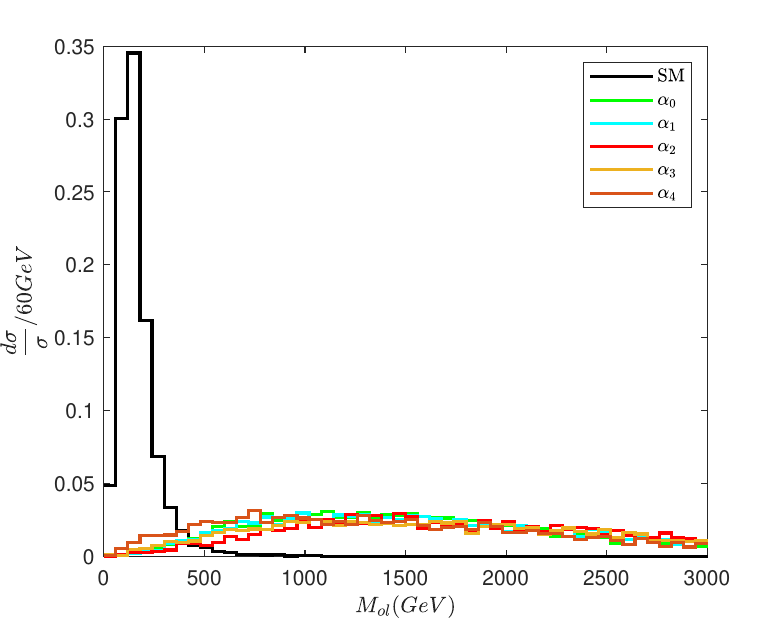}}\hfill
\subfloat[$\cos (\theta _{\ell\ell})$]{\includegraphics[width=0.5\textwidth]{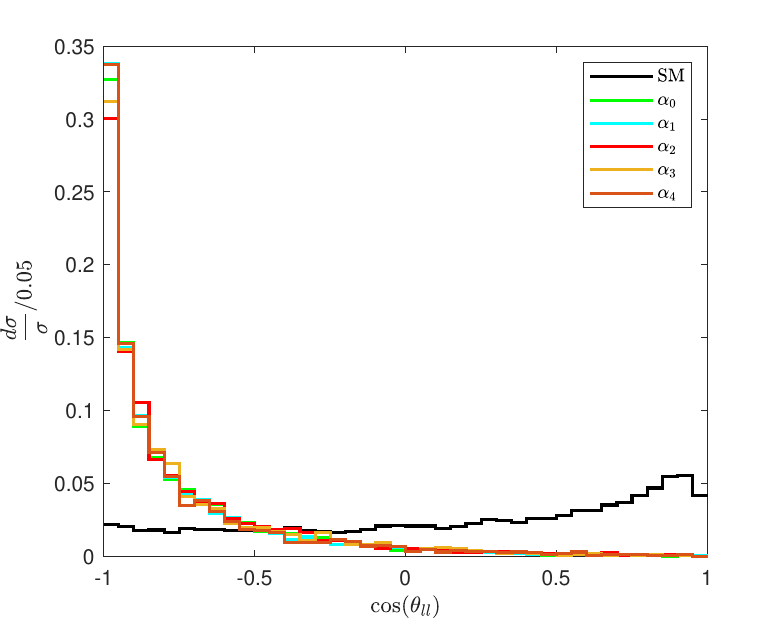}}\vfill
\caption{The differential cross-sections of the SM background and signals as a function of $M_{o1}$ (left panel) and $\cos (\theta _{\ell\ell})$ (right panel) after the standard VBS/VBF cuts.}
\label{fig:cuts}
\end{figure}

The standard VBF/VBS cuts are used to discriminate the VBS from non-VBS process, but we need to also to discriminate aQGC VBS from SM VBS.
It has been studied in the same sign $W$-boson process that $M_{o1}$ is sensitive to anomalous quartic gauge operators except for $O_{S_i}$~\cite{SameSignWWAndMolCut}.
This variable is defined as
\begin{equation}
\begin{split}
&M_{o1}\equiv \sqrt{\left(|\vec{p}^T_{\ell^+}|+|\vec{p}^T_{\ell^-}|+|\vec{p}^{\rm miss}_{\rm T}|\right)^2-\left|\vec{p}^T_{\ell^+}+\vec{p}^T_{\ell^-}+\vec{p}^{\rm miss}_{\rm T}\right|^2},
\end{split}
\label{eq.3.2}
\end{equation}
which provides a very efficient discrimination between signal and backgrounds as shown in Fig.~\ref{fig:cuts}.~(a). We select events with $M_{o1}>500{\rm\ GeV}$.

The $W^{\pm}$ bosons in VBS process should be dominantly back-to-back.
For energetic $W^{\pm}$ bosons, the flight directions of leptons are close to $W^{\pm}$ bosons.
Therefor the leptons should also be dominantly back-to-back, which leads to a small $\cos (\theta _{\ell\ell})$ where $\theta _{\ell \ell}$ is the angle between the leptons.
The differential cross-sections as functions of $\cos (\theta _{\ell\ell})$ are shown in Fig.~\ref{fig:cuts}.~(b), and we choose the cut as $\cos (\theta _{\ell\ell}) < -0.75$.
The efficiency of these cuts are listed in Table.~\ref{tab.cuts}. The basic cuts are from the \verb"MadGraph5_aMC@NLO" toolkit.

\begin{table}
\begin{center}
\begin{tabular}{c|c|c|c|c|c|c}
\hline
$\sigma$($\rm fb$) &$\alpha _0=0.12$ & $\alpha _1=0.2 $ & $\alpha _2=2.9 $ & $\alpha _3=2.3$ & $\alpha _4=5.9$& SM \\
&(TeV$^{-2}$)&(TeV$^{-2}$)&(TeV$^{-4}$)&(TeV$^{-4}$)&(TeV$^{-4}$)& \\
\hline
basic cuts             &$140.8$ &$26.5$ &$28.5$ &$8.8$& $2.0$ & $720.1$ \\
$N_\ell=2$,$N_j\geq 2$             &$80.3$  &$15.0$ &$15.7$ &$4.8$& $1.1$ & $319.3$ \\
$M_{jj}>180{\rm\ GeV},\ \Delta y_{jj}>2.3$              &$60.6$  &$11.6$ &$11.9$ &$3.6$& $0.82$ & $108.1$\\
$M_{o1}>500{\rm\ GeV}$              &$57.9$  &$11.1$ &$11.6$ &$3.5$& $0.74$ & $6.0$  \\
$\cos(\theta _{ll})< -0.75$  &$40.2$  &$8.0$  &$7.9$  &$2.4$& $0.54$ & $0.3$ \\
\hline
\end{tabular}
\end{center}
\caption{\label{tab.cuts} Signal and background cross sections (in fb) with consecutive cuts for the $\ell^+\ell^-jj+\slashed{E}$ final states at $\sqrt{s}$= 13 TeV.}
\end{table}

\subsection{\label{level4.2}Significance of the signal}

In the significance analysis, non-VBS aQGC diagrams (Fig.~\ref{fig:typicalSignal}.~(b)) and all possible interference effects are included. In this case, the total cross-section with one vertex at a time denoted as $\sigma_i$, is approximately a bilinear function of $\alpha _i$. After scanning over the parameter space of $\alpha _i$ in Table~\ref{tab.1}, we can obtain the $\sigma_i$ by fitting.
\begin{equation}
\begin{split}
&\sigma _0= \sigma _{SM}-1.36 ({\rm fb}{\rm TeV}^2) \alpha _0 + 2050  ({\rm fb}{\rm TeV}^4) \alpha _0^2,\\
&\sigma _1= \sigma _{SM}+0.266 ({\rm fb}{\rm TeV}^2) \alpha _1 + 135  ({\rm fb}{\rm TeV}^4) \alpha _1^2,\\
&\sigma _2= \sigma _{SM}+0.0369 ({\rm fb}{\rm TeV}^4) \alpha _2 + 0.613  ({\rm fb}{\rm TeV}^8) \alpha _2^2,\\
&\sigma _3= \sigma _{SM}+0.00883 ({\rm fb}{\rm TeV}^4) \alpha _3 + 0.0355  ({\rm fb}{\rm TeV}^8) \alpha _3^2,\\
&\sigma _4= \sigma _{SM}+0.00770 ({\rm fb}{\rm TeV}^4) \alpha _4 + 0.0529  ({\rm fb}{\rm TeV}^8) \alpha _4^2,\\
\end{split}
\label{eq.4.1}
\end{equation}
where $\sigma _{SM}\approx 0.3{\rm\ fb}$ is the cross section of the SM {background after cuts. The fittings of $\sigma _i$ are shown in Fig.~\ref{fig:fit}.

One can find that $\sigma _i$ are different from the Table.~\ref{tab.cuts}.
With the same coefficients, the cross-sections are $\sigma _0= 29.5\;{\rm fb}$ for $\alpha _0 = 0.12\;({\rm TeV^{-2}})$, $\sigma _1= 5.9\;{\rm fb}$ for $\alpha _1 = 0.2\;({\rm TeV^{-2}})$, $\sigma _2= 5.6\;{\rm fb}$ for $\alpha _2 = 2.9\;({\rm TeV^{-4}})$, $\sigma _3= 1.6\;{\rm fb}$ for $\alpha _3 = 5.9\;({\rm TeV^{-4}})$ and $\sigma _4= 0.61\;{\rm fb}$ for $\alpha _4 = 2.3\;({\rm TeV^{-4}})$, respectively.
Compare with the results of only aQGC VBS process, we find $\sigma _{SM}+\sigma _{aQGC,i}>\sigma _i$, where $\sigma _{aQGC,i}$ is the cross section of VBS process induced by aQGC.
This indicates that the cross-section is reduced by the contributions of the interference and s-channel diagrams (Fig.~\ref{fig:typicalSignal}.~(b)).

\begin{figure}
\includegraphics[width=0.5\textwidth]{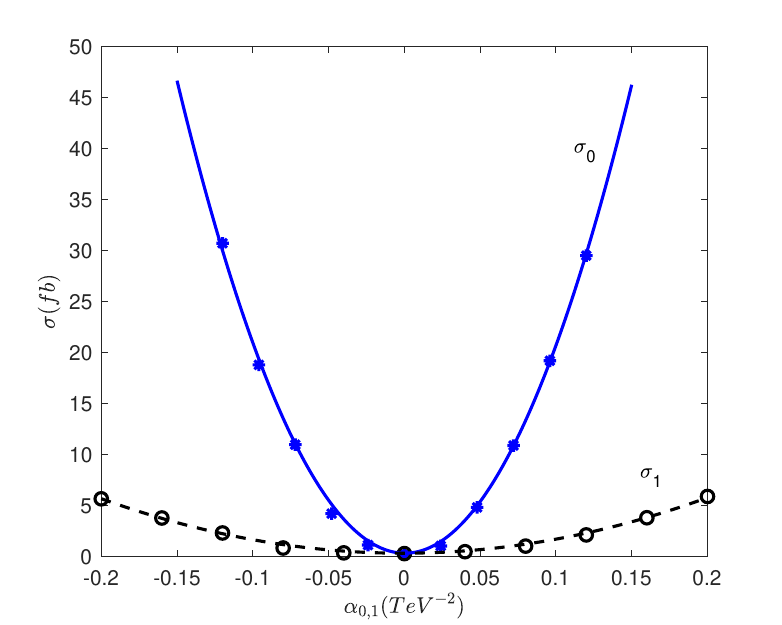}\hfill
\includegraphics[width=0.5\textwidth]{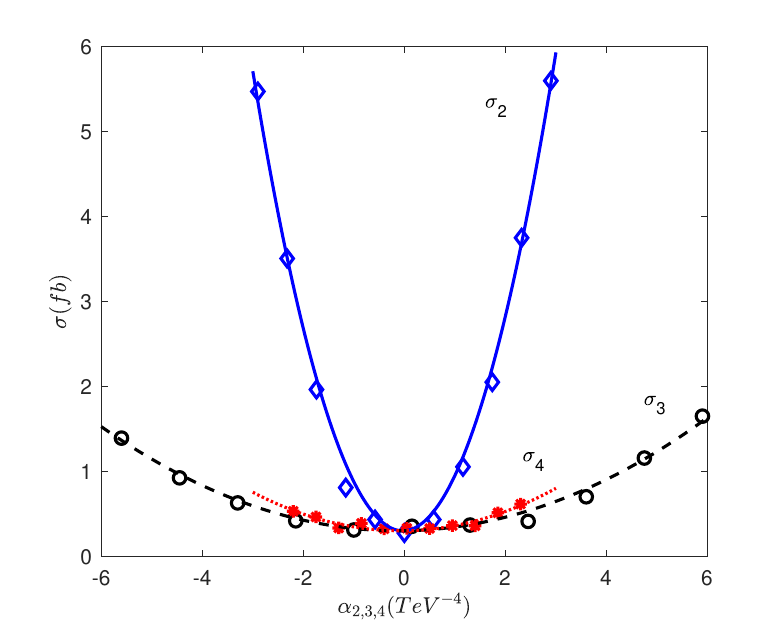}\vfill
\caption{The cross-sections obtained by using $\alpha _{0,1}$ and $\alpha _{2,3,4}$ in the range of Table.~\ref{tab.1}.}
\label{fig:fit}
\end{figure}

We calculate statistical significance (SS) using $SS\equiv N_S/\sqrt{N_S+N_B}$ where $N_S$ and $N_B$ are the numbers of signal and background events, respectively. With SS, we calculate the expected constraints on the vertices and display the results of the constraints in Fig.~\ref{fig:constraints} and Table.~\ref{tab.constraint137} for current LHC luminosity $\mathcal{L}=137.1\;{\rm fb^{-1}}$~\cite{lumino}
\begin{figure}
\subfloat{\includegraphics[width=0.35\textwidth]{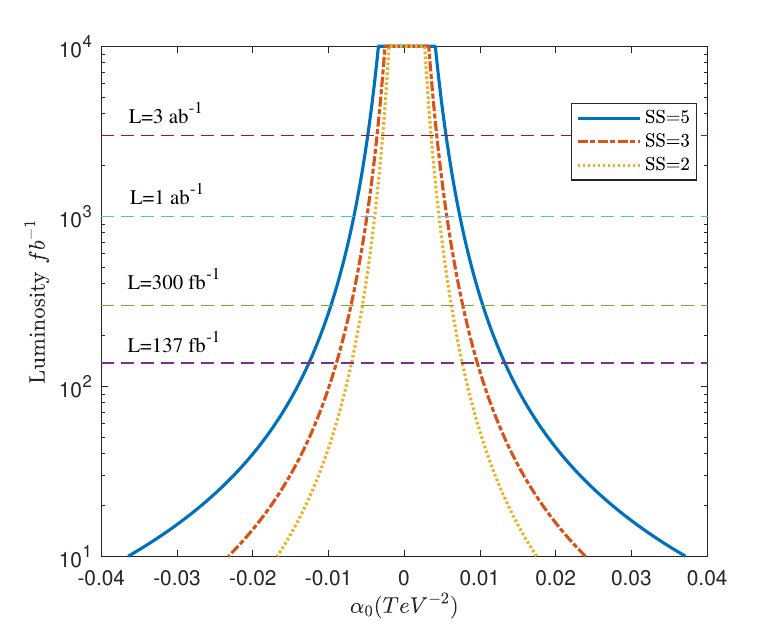}}
\subfloat{\includegraphics[width=0.35\textwidth]{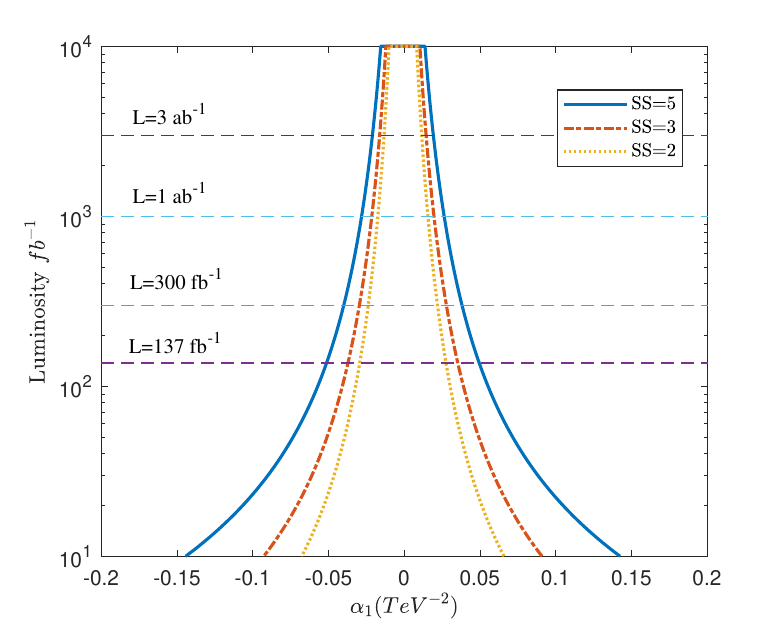}}
\subfloat{\includegraphics[width=0.35\textwidth]{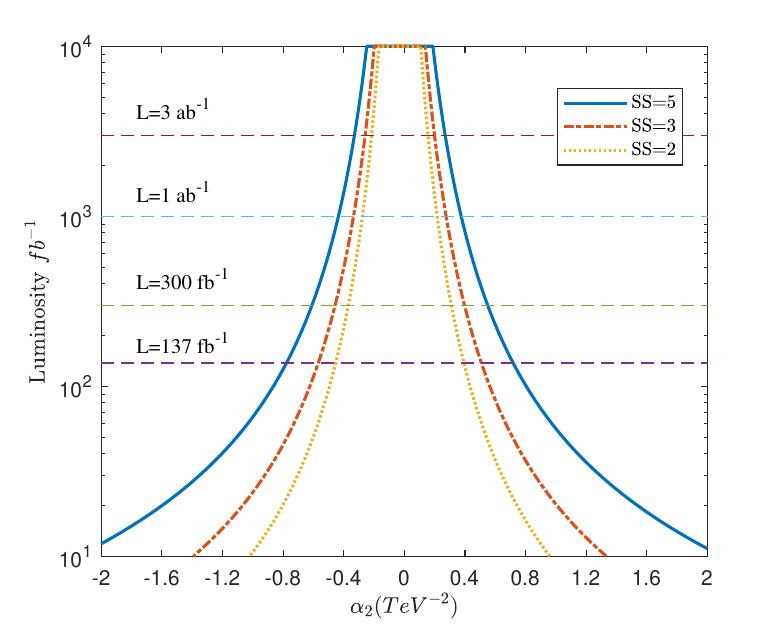}}\vfill
\subfloat{\includegraphics[width=0.35\textwidth]{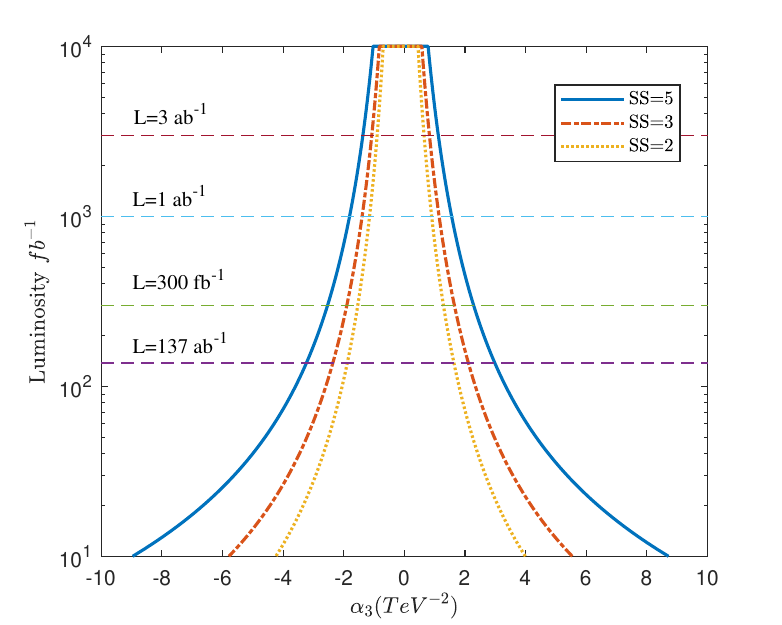}}
\subfloat{\includegraphics[width=0.35\textwidth]{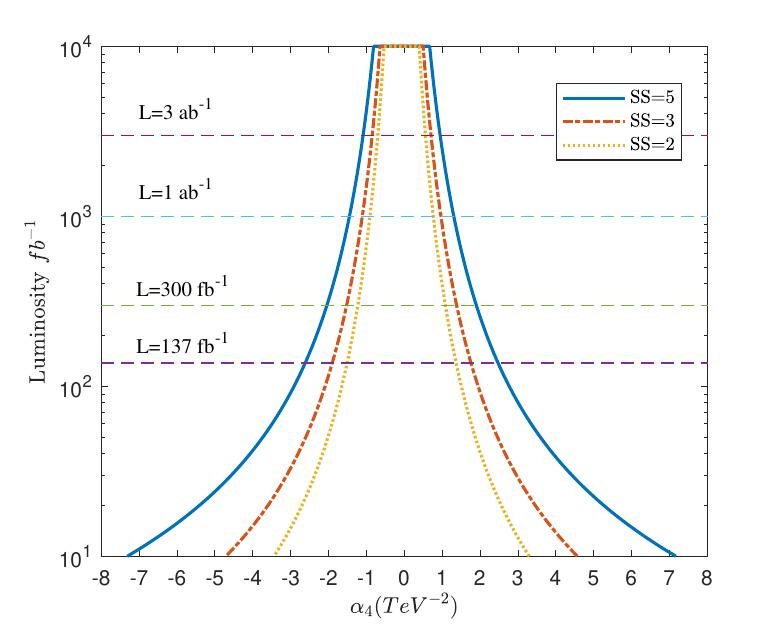}}
\caption{The constraints on the vertices at different luminosities for 2$\sigma$, 3$\sigma$ and 5$\sigma$ statistical significances at $13$ TeV.}
\label{fig:constraints}
\end{figure}

\begin{table}
\begin{center}
\begin{tabular}{c|ccc}
\hline
Constraint & $SS\leq$2 & $SS\leq$3 & $SS\leq$5 \\
\hline
$\alpha_0 ({\rm TeV^{-2}})$ & $[-0.0070, 0.0076]$ & $[-0.0090, 0.0096]$ & $[-0.013, 0.013]$ \\
$\alpha_1 ({\rm TeV^{-2}})$ & $[-0.029, 0.027]$ & $[-0.037, 0.035]$ & $[-0.051, 0.049]$ \\
$\alpha_2 ({\rm TeV^{-4}})$ & $[-0.45, 0.39]$ & $[-0.57, 0.51]$ & $[-0.78, 0.72]$ \\
$\alpha_3 ({\rm TeV^{-4}})$ & $[-1.9, 1.6]$ & $[-2.4, 2.1]$ & $[-3.2, 3.0]$ \\
$\alpha_4 ({\rm TeV^{-4}})$ & $[-1.5, 1.4]$ & $[-1.9, 1.8]$ & $[-2.6, 2.5]$ \\
\hline
\end{tabular}
\end{center}
\caption{\label{tab.constraint137}The constraints on the vertices at $\sqrt{s}=13{\rm\ TeV}$ and at luminosity 137 fb$^{-1}$.}
\end{table}

\section{\label{level5}Summary}

Among the processes measured at LHC, the VBS processes provide excellent opportunities to study the structure of QGCs and possible effects of BSM.
In this letter, we investigate the dimension-8 operators contributing to anomalous $\gamma\gamma WW$ coupling via the VBS process $\gamma\gamma \to W^+W^-$ at the $13$ TeV.
The corresponding $\gamma\gamma WW$ vertices are investigated, the unitarity bounds of those vertices are analyzed.
Our analysis shows that the range of coefficients we picked satisfy the unitarity at $13$ TeV.
To study the observability of the operators, we analyse the signals of aQGCs and backgrounds based on the CMS detector simulation for $jjl^+l^-\nu\bar{\nu}$ final state.
Compared with the SM backgrounds, the $\gamma\gamma WW$ aQGC has unique kinematical features.
We found that the $M_{o1}$ and $\cos (\theta _{\ell\ell})$ are sensitive variables which cut the SM backgrounds efficiently.
For the significance analysis, we take into account the s-channel aQGC effects and the interference between the signal and SM backgrounds.
The contribution of s-channel diagrams induced by the aQGC and the interference effects will decrease the cross section.
Such correction is found to be sizable, and should be considered to ensure an accurate measurement.
Based on the SS calculated, the expected constraints on the $\gamma\gamma WW$ vertices are obtained in Fig.~\ref{fig:constraints} and Table.~\ref{tab.constraint137}.
The $\gamma \gamma \to W^+W^-$ process is found to be sensitive to the $V_{0,AW}$ and $V_{1,AW}$ vertices corresponding to the $O_{M_i}$ operators, and the constraints can be tighten by more than one order of magnitude at 13 TeV LHC with current luminosity.

\section*{ACKNOWLEDGMENT}

\noindent
This work was supported in part by the National Natural Science Foundation of China under Grants No.11905093,  No.11847019 and No.11947066, the Natural Science Foundation of the Liaoning Scientific Committee No.2019-BS-154.

\end{document}